\documentclass[11pt,a4paper]{article}

\usepackage{arxiv}

\usepackage[utf8]{inputenc} 
\usepackage[T1]{fontenc}    
\usepackage{hyperref}       
\usepackage{url}            
\usepackage{booktabs}       
\usepackage{amsfonts}       
\usepackage{nicefrac}       
\usepackage{microtype}      
\usepackage{lipsum}		
\usepackage{graphicx}
\usepackage{natbib}
\usepackage{doi}
\usepackage{amsmath}

\usepackage{setspace}
\title{Plasmonic  photothermal  response of a phantom embedded with  gold nanorod  aggregates  on  broadband  near-infrared irradiation}

\author{{Dheeraj Pratap$^{a,c}$}\thanks{Corresponding author's email: dpratap@iitd.ac.in},  Rizul Gautam$^{a, b}$,  Amit Kumar Shaw$^{a, b}$, Vikas$^{a, b}$, Sanjeev Soni$^{a, b}$ \\ \\
	$^{a}$Biomedical Applications Group, CSIR-Central Scientific Instruments Organization, Sector-30C, Chandigarh-160030, India\\
	$^{b}$Academy of Scientific and Innovative Research (AcSIR), Ghaziabad-201002, India\\ 
	$^{c}$Optics and Photonics Centre, Indian Institute of Technology Delhi, New Delhi-110016, India\\
}

\date{}

\renewcommand{\headeright}{}
\renewcommand{\undertitle}{}

\begin{document}
\maketitle

\begin{abstract}
Longer near-infrared wavelengths provide better penetration depth in biological tissues, so these are useful for plasmonic photothermal cancer therapeutics. In the context of nanoparticles for such applications, the absorption can be tuned for longer NIR wavelengths. However, on increasing the size of the nanoparticle, the scattering is enhanced and thus is not suitable for plasmonic therapeutics. Therefore, to overcome this issue, different types of small gold nanorods were synthesized and converted into stable aggregates to red-shift the plasmonic resonance wavelength to longer near-infrared wavelengths. The gold nanorod aggregates were embedded into the agarose gel phantoms mimicking the tumor-tissue-like structure. The photothermal response was measured through the prepared phantoms using a broadband near-infrared light source. It was shown that even in an extremely dilute concentration of gold nanorods, the photothermal heat generation could increase after the aggregation and also give gives deeper penetration of thermal energy. The observed photothermal response was also verified through numerical simulation. The current study shows better performance by increasing the plasmonic coupling, reducing the mismatch issue of plasmonic resonance shift in the second biological therapeutic window for the aggregates without increasing the size of individual nanoparticles. The aggregates provide better light penetration at deeper tissue by red-shifting the absorbance resonance wavelength which is useful for plasmonic photothermal cancer therapy.  
\end{abstract}


\section{Introduction}
One of the deadliest diseases to affect people in the modern era is cancer. Chemotherapy, radiation, immunotherapy, and surgery are currently the main treatment options but all of these remedies, though, have serious side effects for a patient~\cite{vines2019gold}. For instance, chemotherapy, targets rapidly dividing malignant  as well as the benign cells and causes non-specific damage to the normal cells also due to heterogeneity of cancer~\cite{cobley2010targeting}. Additionally, because malignant cells quickly develop a resistance to chemotherapy, many of the chemo agents lose their effectiveness over time~\cite{lovitt2018doxorubicin}. Contrarily, radiotherapy and immunotherapy can have a detrimental impact on the immune system ~\cite{darby2013risk,postow2018immune}, while surgery, which is frequently used as the first line of treatment for eliminating primary tumors, is invasive and increases the chance of acquiring secondary cancers and infections~\cite{tohme2017surgery}. 

There have been efforts to improve the efficacy and the safety of available therapeutic techniques in order to tackle the complexity of cancer and the risks connected with present treatment options. One of these initiatives is photothermal therapy (PTT), which uses a variety of electromagnetic energy sources, including visible light, infrared (IR), near-infrared (NIR), radio waves, or microwaves, to heat specific regions of tissue in so as to destroy cancer cells through light-to-heat conversion process~\cite{sheng2017review}. PTT has received renewed interest in the recent two decades due to the development of photothermally active nanoparticles, making it a promising method for the non-invasive eradication of malignant tumors~\cite{abadeer2021recent}. There have been reports on several different nanoplatforms, including metal nanoparticles of different compositions, shapes, and architectures, carbon nanoparticles and nanotubes, graphene oxide, and some other inorganic nanoparticles like copper chalcogenide etc.~\cite{yang2019gold,kumar2021recent,kadkhoda2022photothermal}. In the metal nanoparticles, the gold nanoparticles that absorb NIR light, such as hollow gold nanospheres, gold nanostars, gold nanoshells, gold nanorods, and gold nanocages, are the most common type of nanoparticles~\cite{dreaden2012golden,yang2021shape,pakravan2021comparison}. Compared to previously used non-metal materials, these nanoplatforms have higher photothermal conversion characteristics. For deep tissue penetration, the NIR light is advantageous. Additionally, the gold particles have demonstrated high biocompatibility and functioning, offering them a competitive option to other externally delivered medicines that are less effective. As a result, plasmonic photothermal therapy (PPTT), an improved variant of the PTT, is coined~\cite{huang2011plasmonic}. 

The gold nanorods are among the existing NIR-absorbing nanoplatforms that are particularly attractive for PTT applications because of their exceptional characteristics~\cite{xu2019review,pratap2021review,zong2021development}. A variety of gold nanoparticles of different shapes and sizes could be synthesized~\cite{oldenburg1998nanoengineering}.  For example, gold nanorods might be tiny, approximately 25 nm long and 5 nm in diameter~\cite{pratap2022photothermal_a}. On the other hand, the gold nanoshells could have  a diameter greater than 100 nm ~\cite{oldenburg1998nanoengineering}. It is simpler to produce high-quality gold nanorods using the traditional seed-mediated growth method than it is to synthesize gold nanoshells and nanocages~\cite{nikoobakht2003preparation,murphy2005anisotropic}. Gold nanorods display outstanding optical tunability on altering the aspect ratio (AR)~\cite{huang2010gold}. By altering the concentration of silver nitrate used in the seed-mediated growth process, it is possible to change the AR~\cite{granja2021gold}. Apart from this, it is demonstrated that when compared to gold nanoshells, gold nanorods have better photothermal heat generation and blood circulation half-life~\cite{von2009computationally,xie2007quantitative}. Additionally, various bioactive and cancer-curing species can be easily coupled with gold nanorods, paving the way for combinations of therapies like PPTT with chemotherapy or PPTT with photodynamic therapy (PDT). Therefore, multimodal therapy could be possible with the PPTT.

It is well known that for larger size nanoparticles or higher AR of the nanorods, the relative scattering increases . To combat this, numerical results for clusters or aggregates of nanoparticles or nanorods for PPTT applications have recently been reported~\cite{wang2021aggregation,pratap2022photothermal_b}. Experimental evidence was presented for the photothermally stable aggregation of small sized gold nanorods in dispersion form~\cite{pratap2022photothermal_a}. However, the gold nanorod aggregation is rarely investigated for PPTT based studies. In this paper, we report on photothermal properties through tissue phantoms by employing aggregates of gold nanorods to measure the photothermal response experimentally. Instead of using a monochromatic laser in the current investigation, we use broadband NIR light.

In this paper, experimentally we report the photothermal response using different types of gold nanorods aggregates in the agarose gel tissue like phantoms on irradiation with an inhouse developed broadband NIR light source. The organization of the paper is as follows. First, we introduce the problem in the introduction section. The second section describes the synthesis, optical and structural characterization of the nanorods and their aggregates. The photothermal setup and the photothermal response are discussed in the third section. In the fourth section, we show the results and discussions. The last section concludes the paper. 

\section{Synthesis and characterization of nanorods and aggregates}
To synthesize the gold nanorods we followed the standardized protocols described in the literature with slight modifications. For the current study, we made two types of gold nanorods with seed-mediated method-1 and method-2. It should be emphasized that throughout the experiment, the concentration of the gold nanorods synthesized by each method allowed for an equal number of gold nanorods to exist in monodispersive form and in their respective aggregates.  We received ascorbic acid and Bovine Serum Albumin (BSA) from Sisco Research Laboratories, hydrochloric acid (HCl) from Finar, chloroauric acid (HAuCl$_4$), silver nitrate (AgNO$_3$), cetyltrimethylammonium bromide (CTAB),  sodium borohydride (NaBH$_4$) and Dulbecco's Modified Eagle Medium (DMEM) from Sigma-Aldrich for the synthesis of gold nanorods. 
\subsection{Synthesis of smaller nanorods by method-1}
For the first case using method-1, the seed-mediated technique, slightly modified from the process described in references~\cite{jia2015synthesis,chang2018mini}, was successful in producing the necessary gold nanorods. This protocol also demonstrated the reproducibility of synthesized gold nanorods.  A 2~mL (0.01~M) solution of NaBH$_4$ was prepared and stored in the freezer for use as ice-cold in the first step. Under vigorous stirring, 0.25~mL (0.01~M) of HAuCl$_4$ solution was added to 9.75~mL (0.1~M) CTAB solution. After adding 0.6~mL (0.01~M) of freshly made, ice-cold NaBH$_4$, the combination of HAuCl$_4$ and CTAB was vigorously stirred for two minutes. Before use, the resulting solution that had turned yellowish-brown was held at room temperature for two hours. This served as the starting point as a seed solution for the creation of nanorods. The growth solution was made by adding 5~mL (0.01~M) HAuCl$_4$, 1~mL (0.01~M) AgNO$_3$, 2~mL (1.0~M) HCl sequentially into the 90~mL (0.1~M) CTAB solution, and then adding 0.8~mL (0.1~M) fresh ascorbic acid solution while rapidly stirring until the solution became transparent. To grow the nanorods, the entire amount of seed solution was added to the growth solution and vigorously stirred for two minutes. The finished mixture was then left overnight undisturbed at 35~$^{\circ}$C. The following day, using a Remi-C24 plus centrifuge, the entire solution containing the nanorods was centrifuged by separating it into equal portions and placing them in four 50~mL round-bottom centrifuge tubes. This was done at 15000~rpm at 30~$^{\circ}$C for 20 minutes. The extra CTAB was gathered as a supernatant and subsequently thrown away. Following the second wash in DI water under identical conditions, the synthesized gold nanorods formed in the pellet were disseminated in water. The amount of pellet and added DI water in each step of the washing process were carefully recorded to calculate the concentration of CTAB empirical in the final nanorod solution as reported previously~\cite{wu2015large,pratap2022photothermal_a}. The total volume of the final solution was 24~mL and kept as stock suspension of nanorods for further uses which had a measured concentration of 15.9 mg/L using the inductively coupled plasma mass spectrometry (ICP-MS). We abbreviated this stock gold nanorod as sNR$_1$. 
\subsection{Synthesis of larger nanorods by method-2}
In method-2, gold nanorods were synthesized using the seed-mediated method but had a higher localized surface plasmon resonance (LSPR) wavelength than the above method-1. In this case, also, the seed solution and the growth solution were prepared. Firstly, seed solution was prepared by stirring 10~mL (0.1~M) CTAB solution with 0.25~mL (0.01~M) HAuCl$_4$ followed by addition of freshly prepared ice-cold solution of 0.6 mL (0.01~M) NaBH$_4$. The ice-cold NaBH$_4$ was added quickly to the stirring mixture of CTAB and HAuCl$_4$ which immediately turned the bright yellow colour of the solution into a yellowish brown. This colour change of the solution confirms the formation of gold seeds. The obtained gold seeds solution was stirred for another 10 minutes followed by incubating the solution at $27~^{\circ}$C for 2.5 hours. After completion of the incubation period, the growth solution was prepared by mixing freshly prepared 80~mL (0.1~M) CTAB with 4~mL (0.01~M) HAuCl$_4$. Further a 0.4~mL (1~N) HCl was added to the above solution. To this solution, 0.8~mL (0.01~M) AgNO$_3$ was added under stirring conditions followed by drop-wise addition of 0.6~mL (0.1~M) ascorbic acid to the reaction mixture that tuned the coloured solution into colourless. To this colourless solution, $192~\mu$L of incubated seed solution was introduced and again the solution was incubated at $27~^{\circ}$C for the next 20 hours. After completion of the incubation period, the obtained gold nanorods solution was centrifuged at 15000~rpm (Remi-C24 plus centrifuge) for 15 minutes as the synthesized gold nanorods were equally divided into 2~mL eppendorf tube (all 18 eppendorfs tubes). The excess amount of CTAB collected in the form of supernatant after centrifugation was collected and discarded. The obtained gold nanorods in the form of pellets were again washed in DI water at the same centrifugation condition. The CTAB concentration in the final gold nanorods suspension solution was calculated again as discussed in previous reports~\cite{wu2015large,pratap2022photothermal_a}. The total volume of the final solution was 5~mL and stored at $4~^{\circ}$C for further experimental work. The final concentration of gold nanorods in the suspension solution was examined/measured using ICP-MS with a concentration of 663~mg/L. The abbreviation of this set of the stock gold nanorods was sNR$_2$.
\subsection{Conversion of nanorods into stable aggregates}
For the PPTT study and to further dilute the gold nanorod's stock suspensions (sNR$_1$ and sNR$_2$) and to ensure that there was no plasmonic coupling among the nanorods, a 950~$\mu$l of DI water and 50~$\mu$l of stock suspension were combined to create a further 1:19 diluted samples in the monodispersive forms NR$_1$ and NR$_2$ as obtained from synthesis method-1 and method-2. The diluted nanorods suspensions had a very subtle pink tint to them. As a  reference for the aggregates to be created, the suspension of diluted gold nanorods was used for comparison. In the next step, we used DMEM and BSA to convert synthesized gold nanorods into aggregates. By dissolving BSA in DMEM, we created a solution that was 4~\% (w/v). For the aggregation, a 50~$\mu$l of stock suspension of the sNR$_1$n was added into a 950~mL (4\%) BSA-DMEM solution and mixture was vortexed at 1900~rpm for five minutes under ambient conditions. Apart from the reference monodispersive sample, NR$_1$, now we had a sample of aggregates, NR$_1$+4\%BSA, corresponding to 4\% BSA-DMEM aggregation. The same former aggregation procedure had been repeated for sNR$_2$ to convert into aggregates and its samples was named as NR$_2$+4\%BSA.  
\section{Structural characterization}
\begin{figure}[h]
\centering
\includegraphics[width=0.8\textwidth]{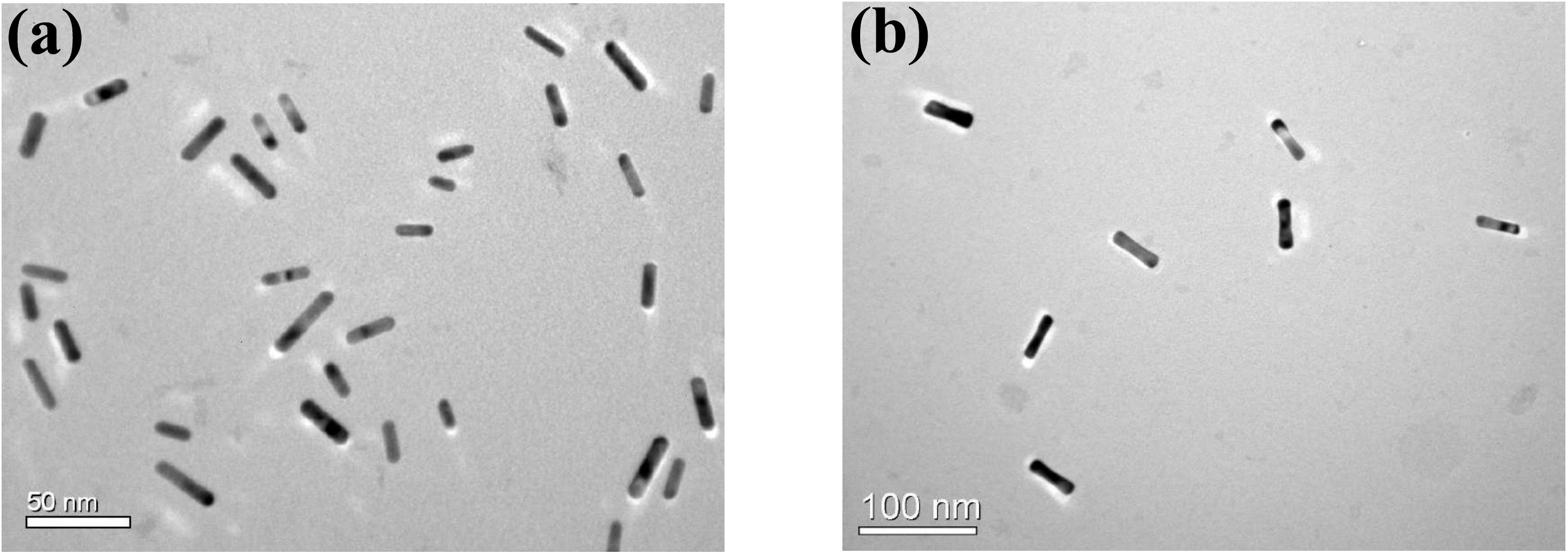}
\caption{Scanning electron microscope (TEM) images of gold nanorods prepared by (a) method-1 and (b) method-2. The scale bar in (a) is 50~nm and in (b) is 100~nm.}
\label{fig:tem_images}
\end{figure}
\begin{figure}[h]
\centering
\includegraphics[width=0.8\textwidth]{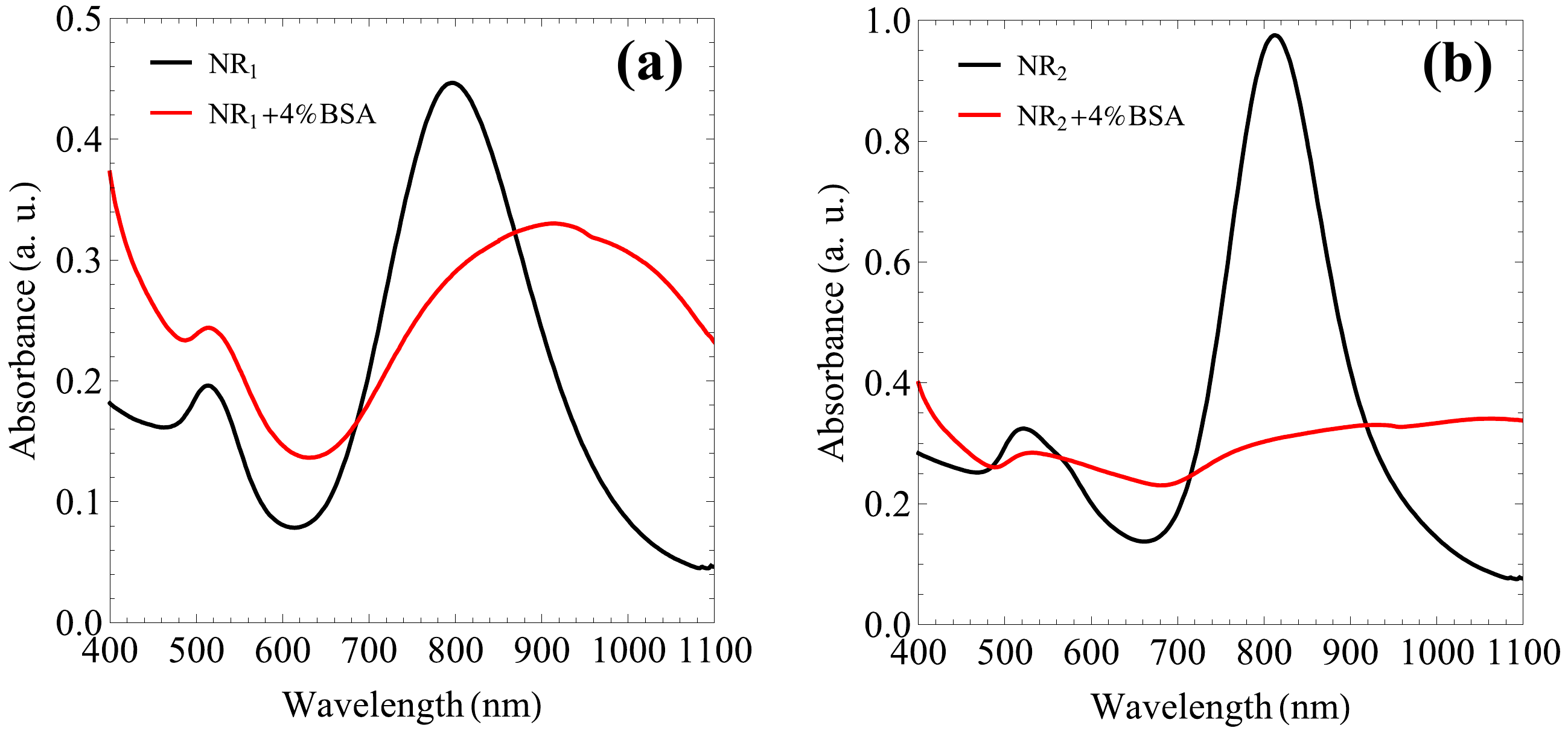}
\caption{Absorption spectra of (a) NR$_1$ and aggregates NR$_1$+4\%BSA, and (b) NR$_2$ and aggregates NR$_2$+4\%BSA.}
\label{fig:NR1_NR2_abs}
\end{figure}

 
The structural characterization of gold nanorods was carried out with a transmission electron microscope (TEM) of make JEM-2100CR, JEOL. Figure~\ref{fig:tem_images} shows the TEM images of the NR$_1$ and NR$_2$. The average measured size of the NR$_1$ was 24.6 $\pm$ 3.3 nm $\times$ 5.9 $\pm$ 0.1 nm and of the NR$_2$ was 42.4 $\pm$ 2.9 nm $\times$ 9.4 $\pm$ 0.7 nm. We see that the size of later gold nanorods was roughly two times larger than the former one. Since the stock solutions of the nanorods were extremely diluted, therefore, the nanorods are sparsely distributed in the images. The absorption spectra of the monodispersive gold nanorods and their aggregates were measured using a UV-visible spectrometer (UV 3200, LABINDIA). Figure~\ref{fig:NR1_NR2_abs} shows the absorption spectra of the gold nanorods and their aggregates. The strongest LSPR of the NR$_1$ and NR$_1$+4\%BSA were at 797~nm and 915~nm respectively. While the strongest LSPR for the NR$_2$ and NR$_2$+4\%BSA were at 812~nm and 1062~nm respectively. The strongest LSPR wavelengths of the nanorods and aggregates are listed in Table~\ref{tab:SPR_and_dT}. It could be seen that when BSA was 4\% then a significant the LSPR red-shift was observed. The absorption spectra of aggregates associated with NR$_2$ are much flatter than that of the individual nanorods. However, as reported previously, these were optically and photothermally stable~\cite{pratap2022photothermal_a}. It is clear from Fig.~2 that by properly choosing the size of the nanorods and the concentration of BSA, we can easily tune the LSPR from the first biological therapeutic window to the second biological therapeutic window. It should be noted that any other nanoparticles of different shapes also could be employed for the higher red shifting of the LSPR. 
\begin{table}[h]
\centering
\caption{The LSPR wavelength of the gold nanorods and aggregates, and the maximum temperature rise $\Delta$T~($^{\circ}$C) after 15 minutes of irradiation of the agarose gel phantoms that were recorded with the thermocouples 1, 2, 3 and 4 at the depth of 1~mm, 5~mm, 10~mm and 15~mm respectively. }
\begin{tabular}{| l | c | c | c | c | c |}
\hline 
Sample & LSPR (nm) & $\Delta$T$_1$ ($^{\circ}$C) & $\Delta$T$_2$ ($^{\circ}$C) & $\Delta$T$_3$ ($^{\circ}$C) & $\Delta$T$_4$ ($^{\circ}$C)\\ 
\hline \hline
NR$_1$        & 797  & 19.67 & 17.74 	&   12.18	&  7.40 \\ \hline  
NR$_1$+4\%BSA & 915  & 23.57 & 19.66	&   15.24	&  8.05 \\ \hline  
NR$_2$        & 812  & 21.67 & 17.90 	&   13.92	&  8.46 \\ \hline  
NR$_2$+4\%BSA & 1062 & 20.76 & 15.68	&   12.57	&  8.23 \\ \hline  
\end{tabular}
\label{tab:SPR_and_dT}
\end{table}
\section{Photothermal set-up and agarose gel phantom}
\begin{figure}[h]
\centering
\includegraphics[width=1\textwidth]{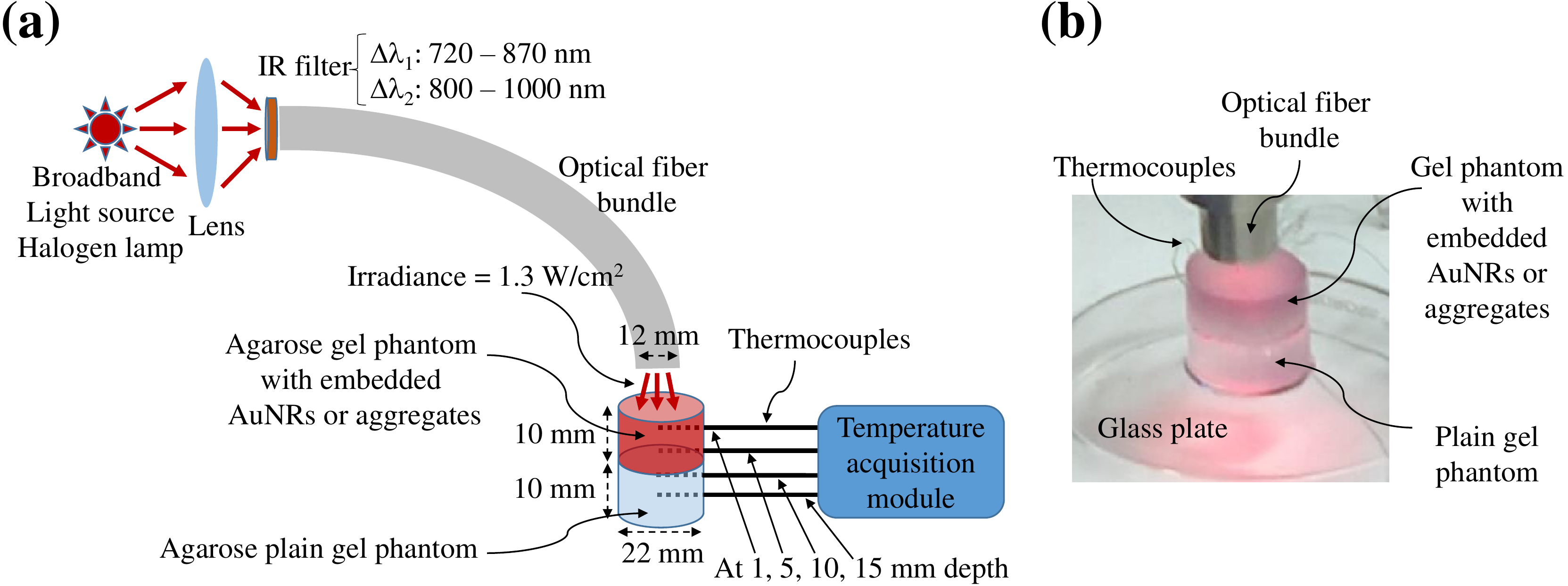}
\caption{(a) Photothermal set-up and measurement plan, and (b) prepared agarose gel phantom (1\%) embedded with gold nanorods/aggregates on the top region kept on a glass plate.}
\label{fig:exp_plan_and_gel_phantom} 
\end{figure}

The photothermal experiment was carried out using in-house developed NIR broadband light source. . Figure~\ref{fig:exp_plan_and_gel_phantom}(a) shows the experimental plan of the PPTT measurement. A broadband NIR light source was used to irradiate the phantom. Since the nanorods and aggregates have different LSPR wavelengths that lie in a wide range of the light spectrum, therefore, two different optical bandpass filters have been used to cut off the extra wavelength bands of the radiation. The first filter-1 has a transmission range of 720~nm - 870~nm ($\Delta \lambda_1$) and the second filter-2 has a transmission range of 800~nm - 1000~nm ($\Delta \lambda_2$). The temperature for the first and the second phantoms as listed in Table~\ref{tab:SPR_and_dT} were irradiated using filter-1, the rest of all four samples (aggregated ones) were irradiated using filter-2. A similar photothermal measurement with braodband visible light (filter) had been reported by Soni et al.~\cite{soni2015experimental}.  

The transmitted light after the infrared filter was collected through an optical fiber bundle of diameter 12~mm and irradiated to tissue gel phantom. The incident intensity of light at the top of the phantom was 1.3~W/cm$^2$. The total incident optical power after the fiber bundle was 1.8~W. Utilizing "K" type thermocouples, the temperature of the gel phantom was determined at the depths of 1~mm, 5~mm, 10~mm, and 15~mm from the top surface of the phantom region and named probes 1, 2, 3, and 4 respectively. Using a data acquisition equipment from National Instruments  (NI-myDAQ), the thermocouple's output was recorded. On irradiation of the phantom for 15 minutes, the temperature change for heating was recorded during the photothermal study and cooling for the same time duration, i.e., 15 minutes after turning off the irradiation. 

The gel phantom was prepared with agarose powder (1\%)  by dissolving it into the DI water or nanorods/aggregates suspension considered for the photothermal experiment. The shape of the agarose gel phantom was cylindrical. Its diameter was 22~mm and total height was 20~mm as shown in the schematic Fig.~\ref{fig:exp_plan_and_gel_phantom}(a) and the actual photograph in Fig.~\ref{fig:exp_plan_and_gel_phantom}(b). The upper part of the gel of height of 10~mm was embedded with the gold nanorods or aggregates and the lower part of the height of 10~mm was plain gel. The top part of the phantom embedded with gold nanorods or aggregates represents the tumor region and the lower plain part represents the healthy tissue. The irradiation was on the top surface of the agarose gel phantom and for support, the whole gel phantom was kept on a plain glass plate.  
\section{Results and discussions}
\begin{figure}[h]
\centering
\includegraphics[width=1\textwidth]{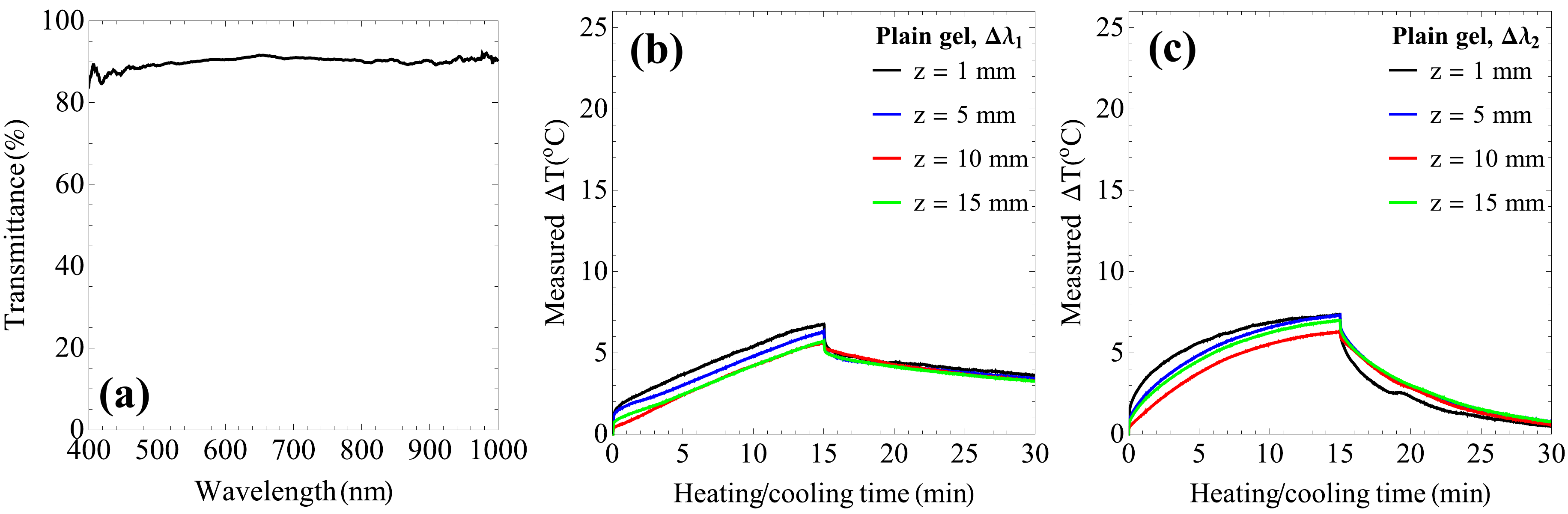}
\caption{(a) Transmission of the glass plate used in the photothermal experiment. Temperature rise $\Delta T$ ($^{\circ}$C) in the plain agarose gel (1\%) irradiated with NIR bandpass (b) filter-1 having the transmission $\Delta \lambda_1 = 720 - 870 $ nm  and (c) filter-2 having the transmission $\Delta \lambda_2 = 800 - 1000 $ nm. }
\label{fig:glass_trans_plain_gel_heating}
\end{figure}

The transmission of the glass plate is shown in Fig.~\ref{fig:glass_trans_plain_gel_heating}(a). The average transmission of the plate was more than 90~\%. Prior to irradiation of the phantom embedded with gold nanorods or aggregates, the plain gel of total height 20~mm was illuminated for 15 minutes to see the effect of water heating in the gel phantom. The temperature rise ($\Delta T$) using infrared bandpass filter-1 and filter-2 are shown in Fig.~\ref{fig:glass_trans_plain_gel_heating}(b) and Fig.~\ref{fig:glass_trans_plain_gel_heating}(c) respectively. At the end of illumination with filter-1 and filter-2 after exposure for 15 minutes, the maximum difference in the temperature rise was 0.5~$^{\circ}$C in the probe-1 at the depth of 1~mm. So, the two different NIR filters can be used to cut off the undesired light spectrum to exclusively evaluate the photothermal response of the GNR and its aggregates.  

\begin{figure}[h]
\centering
\includegraphics[width=1\textwidth]{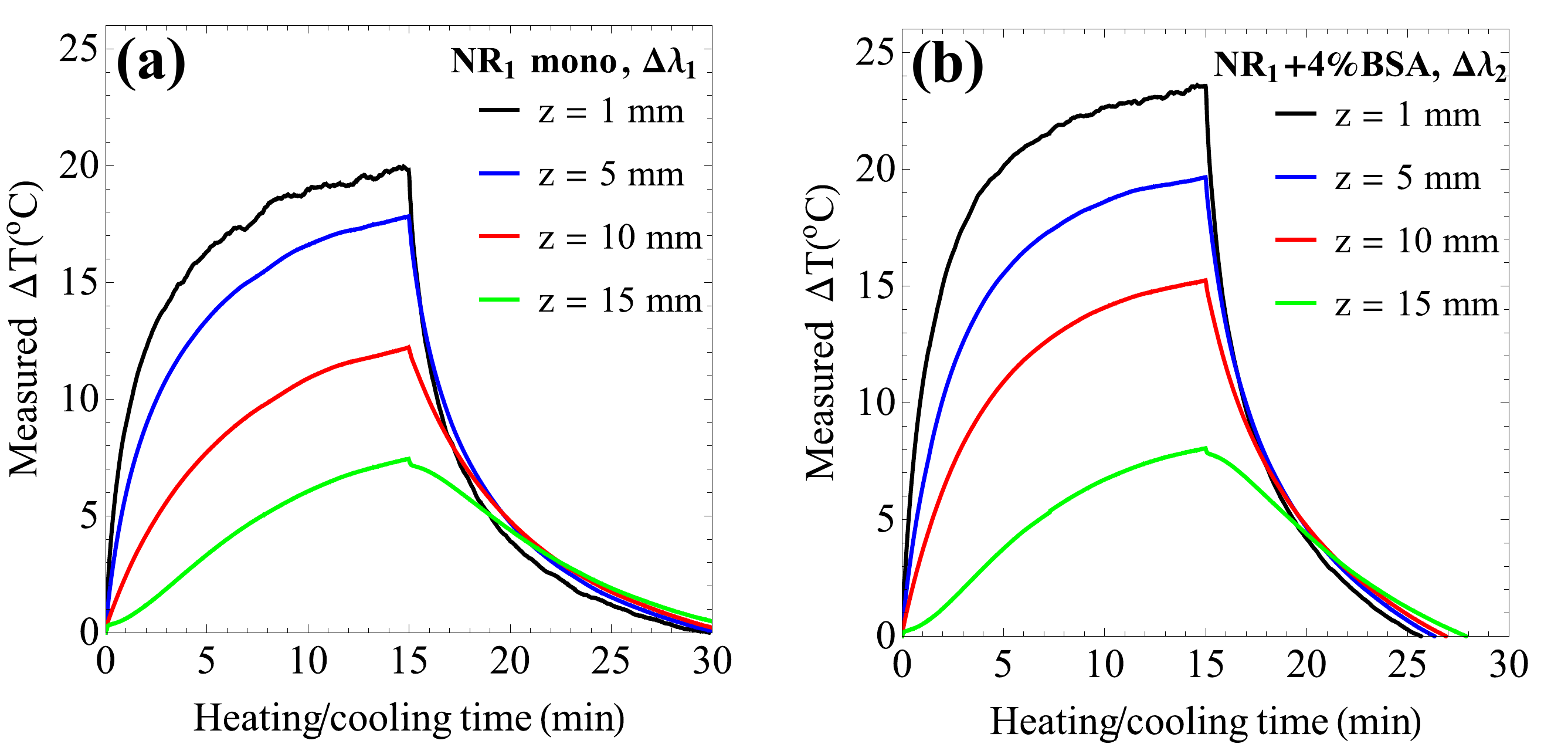}
\caption{Temperature rise $\Delta T$ ($^{\circ}$C) in the agarose gel phantom embedded with (a) gold nanorods prepared by method-1, and its (b) aggregates with 4\% BSA. Sample NR$_1$ in (a) was irradiated with spectral  range-1, $\Delta \lambda_1 = 720 - 870 $ nm, and sample NR$_1$+4\%BSA in (b) irradiated with spectral  range-2, $\Delta \lambda_2 = 800 - 1000 $ nm.}
\label{fig:NR1_pptt}
\end{figure}

The temperature rise in the gel phantom embedded with gold NR$_1$ and their aggregates are shown in Fig.~\ref{fig:NR1_pptt}. The maximum temperature change for the monodispersive case was 19.67~$^{\circ}$C for the probe-1 at a depth of 1~mm. There was temperature change of 23.57~$^{\circ}$C in the case of aggregates with 4\% BSA that showed the strong plasmonic coupling among the gold nanorods. The difference in temperature rise between samples NR$_1$ and NR$_1$+4\%BSA is 3.90~$^{\circ}$C. We see that the strongest LSPR wavelength 915~nm  of sample NR$_1$+4\%BSA lies in the transmission range of infrared bandpass filter-1. Thus it is better to use an aggregated form of the nanoparticles for the therapeutic application to get more temperature rise in the same concentration of the original monodispersive nanoparticles. The maximum change in temperature measured for the NR$_1$ and their aggregates NR$_1$+4\%BSA with probes 1 , 2, 3 and 4 at depth 1~mm, 5~mm, 10~mm and 15~mm, respectively, are listed in Table~\ref{tab:SPR_and_dT}. 
\begin{figure}[h]
\centering
\includegraphics[width=1\textwidth]{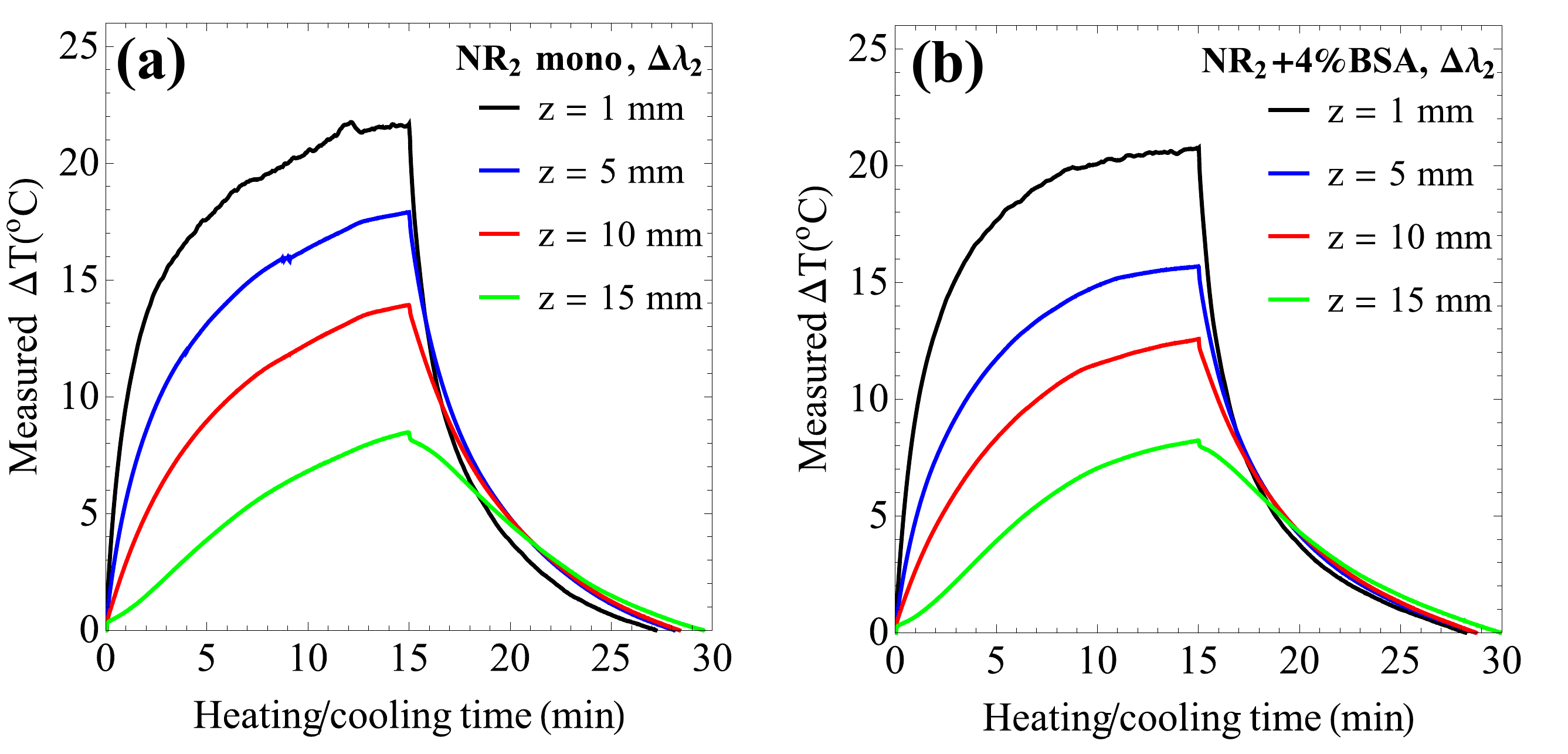}
\caption{Temperature rise $\Delta T$ ($^{\circ}$C) in the tissue gel phantom embedded with (a) gold nanorods prepared by method-2, and its (b) aggregates with 4\% BSA. Samples were irradiated with spectral range-2, $\Delta \lambda_2 = 800 - 1000 $ nm.}
\label{fig:NR2_pptt}
\end{figure} 
The temperature rise in the case of the gold NR$_2$ and their aggregates NR$_2$+4\%BSA embedded in the gel phantom is quite similar to the former case of the nanorods. Figure~\ref{fig:NR2_pptt} shows the temperature rise of the latter case. The maximum temperature rise was approximately 21.67~$^{\circ}$C for the monodispersive case for probe-1 at the depth of 1~mm. For the aggregates with 4\%BSA of the maximum temperature change was 20.76~$^{\circ}$C respectively. Here we see that the temperature is decreased. This is due to the reason that the absorbance was significantly less than the constitutive monodispersive form of the nanorods and its strongest LSPR peak was out of the wavelength range of the bandpass filter-2. The difference in the temperature rise between NR$_2$ and NR$_2$+4\%BSA is 0.91~~$^{\circ}$C. However, due to the flatness of the absorbance of aggregated nanorods and the use of a broadband light source, there was still a sufficient temperature rise for thermal ablation purpose. Therefore, it would be better if we use the standardized broadband light source concerning the nanoparticles for the PPTT applications. 

Comparing Fig.~\ref{fig:NR1_pptt} and Fig.~\ref{fig:NR2_pptt}, we notice that as the depth increases, the rise in temperature decreases as expected. In both cases, the temperature rise trends are almost similar. However, there were differences in the temperature rise due to the different sizes and, shapes of the nanorods synthesized by different methods. Additionally, the amplitudes of absorbances (a.u.) and the nanorods sharpness (or flatness) were also responsible for further temperature rises in both cases. A third possible reason for the lower temperature rise in the 4\% aggregates of NR$_2$ is the occurrence of absorbance (LSPR = 1062~nm) beyond the spectral range of filter-2, i.e., $\Delta \lambda_2 = 800 - 1000 $ nm. If its LSPR had been in the spectral range of filter-2 then there might be a higher temperature change compared to its monodispersive form, i.e., NR$_2$, as observed in the case for aggregates of NR$_1$. 

\begin{figure}
\centering
\includegraphics[width=1\textwidth]{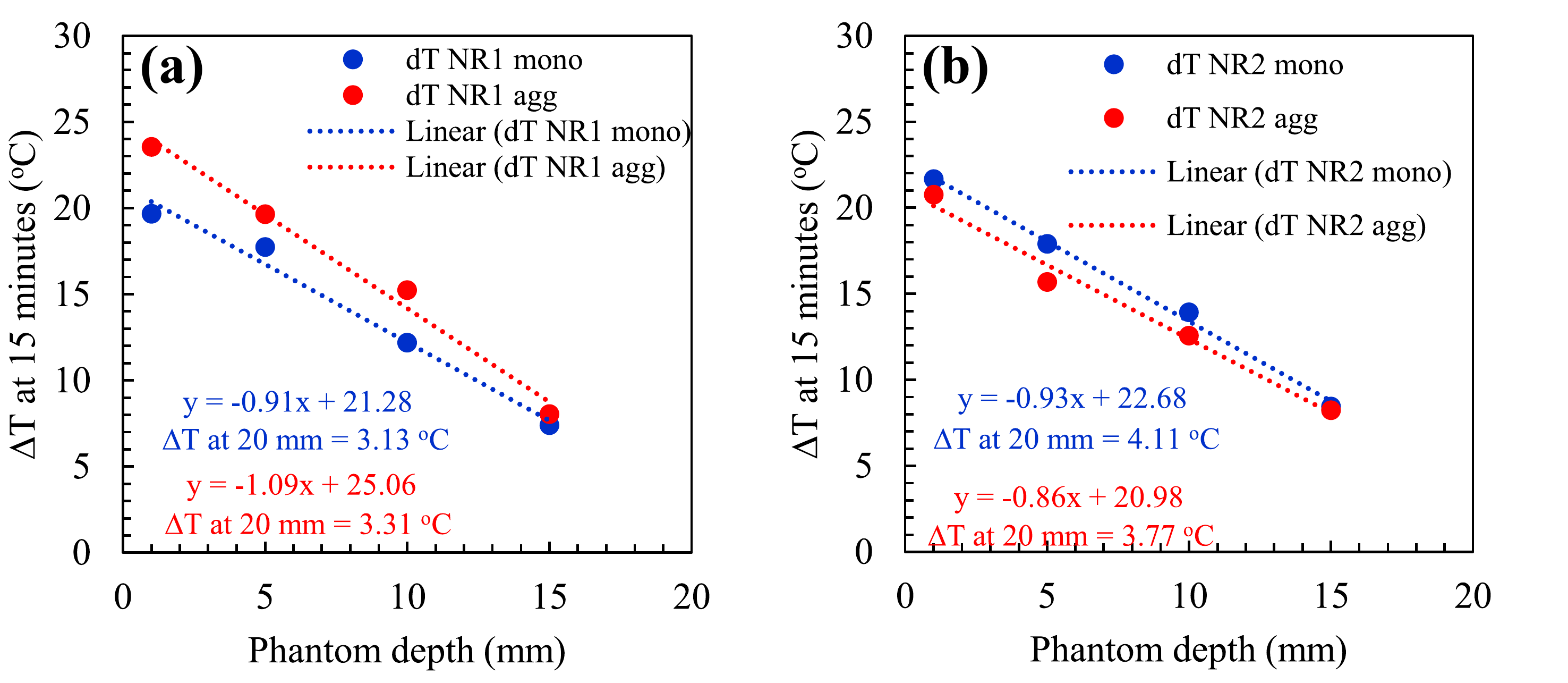}
\caption{Linear fitting of temperature rises $\Delta T$ ($^{\circ}$C) at 15 minutes in the agarose gel phantom embedded with gold nanorods and their aggregates prepared by (a) method-1, and (b) method-2. The extrapolated temperature rise at depth of 20~mm is shown in the inset.}
\label{fig:dT_at_20mm}
\end{figure}
As for longer NIR wavelengths, the light has deeper tissue penetration. Higher penetration depth with aggregation was also confirmed with the experimental data. To do so, we plotted the temperature rise after 15 minutes of irradiation concerning the depth of the thermal probe for method-1 (blue line) and method-2 (red line) set of samples in Fig.~\ref{fig:dT_at_20mm}(a) and Fig.~\ref{fig:dT_at_20mm}(b) respectively. The experimental data were linearly fitted and their corresponding equations are shown inset of the figures. Using the equations of the linear fitting, we calculated the temperature rise at the depth of 20~mm. For the first set of sample, temperature rise for monodispersive (NR$_1$) is 3.13~$^{\circ}$C and for aggregate (NR$_1$+4\%BSA) is 3.31~$^{\circ}$C. For the second set of samples, the temperature rise at 20~mm depth for monodispersive (NR$_2$) is 4.11~$^{\circ}$C and for aggregate (NR$_2$+4\%) is 3.77~$^{\circ}$C.  We can see that for the first set of samples, due to the aggregation there is a higher temperature rise at the depth of 20~mm while for the second set of samples it is not. This is due to the well-defined resonance peak that lies within the $\Delta \lambda_2$ spectral range while the resonance peak of the second set of aggregates is out of the spectral range of incident radiation. Although the temperature difference at the 20~mm depth between the monodispersive and aggregate cases is small, 0.18~$^{\circ}$C only, it proves that the higher temperature rise at the deeper phantom (or tissue) is possible due to the aggregation of the nanoparticles. Therefore, the aggregates compared to monodispersive could be a better option for PPTT.

\section{Computational Modeling}
\begin{figure}[h]
\centering
\includegraphics[width=1\textwidth]{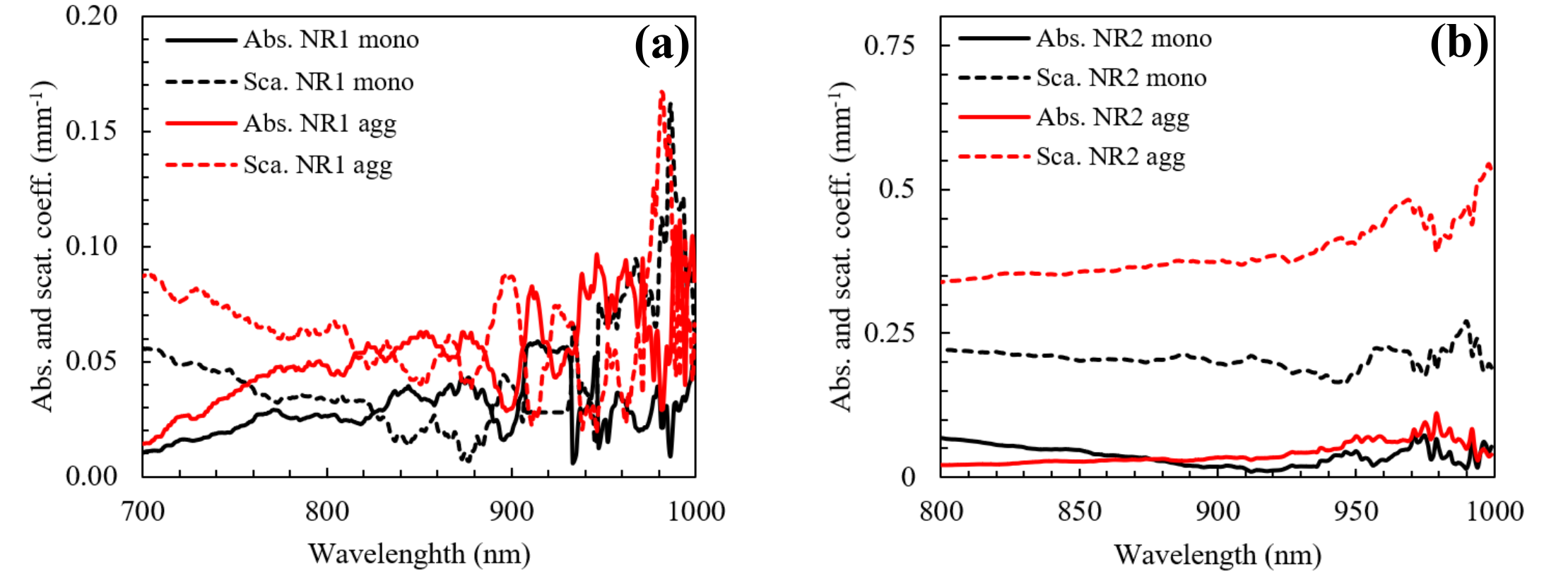}
\caption{Absorption and scattering coefficients of the agarose gel phantom embedded with (a) NR1 and aggregates and (b) NR2 and aggregates, as obtained by the inverse-adding-doubling method (IADM).}
\label{fig:abs_sca_coeff}
\end{figure}

As the nanorods and aggregates are randomly distributed in the gel phantom and it is difficult to homogenize the domain to get the optical parameters. Therefore, to get the optical parameters of the gel phantoms, we used the spectroscopic method. A thin slice of the gel phantom of thickness 3~mm was taken and its reflectance and transmittance spectrum was recorded with an integrating sphere. Using the standard algorithm of the inverse-adding-doubling method(IADM)~\cite{prahl1993determining}, the absorption and scattering coefficients were calculated as shown in Fig.~\ref{fig:abs_sca_coeff}. The size of the nanorods for method-2 is roughly two times larger than the nanorods prepared by method-1. Figure~\ref{fig:abs_sca_coeff}(a) shows that there is higher scattering for smaller wavelengths while it becomes equivalent to absorption for larger wavelengths. In Fig.~\ref{fig:abs_sca_coeff}(b), the scattering is always dominating over the whole wavelength range 700nm-1000nm due to the larger size of the nanorods. The coefficients around 1000nm are quite noisy due to the end of the detectors' range.  The calculated coefficients were used to simulate the photothermal effect in the agarose gel phantom embedded with gold nanorods (NR$_1$, NR$_2$) and their aggregates (NR$_1$+4\%BSA, NR$_2$+4\%BSA).

\begin{figure}[h]
\centering
\includegraphics[width=0.4\textwidth]{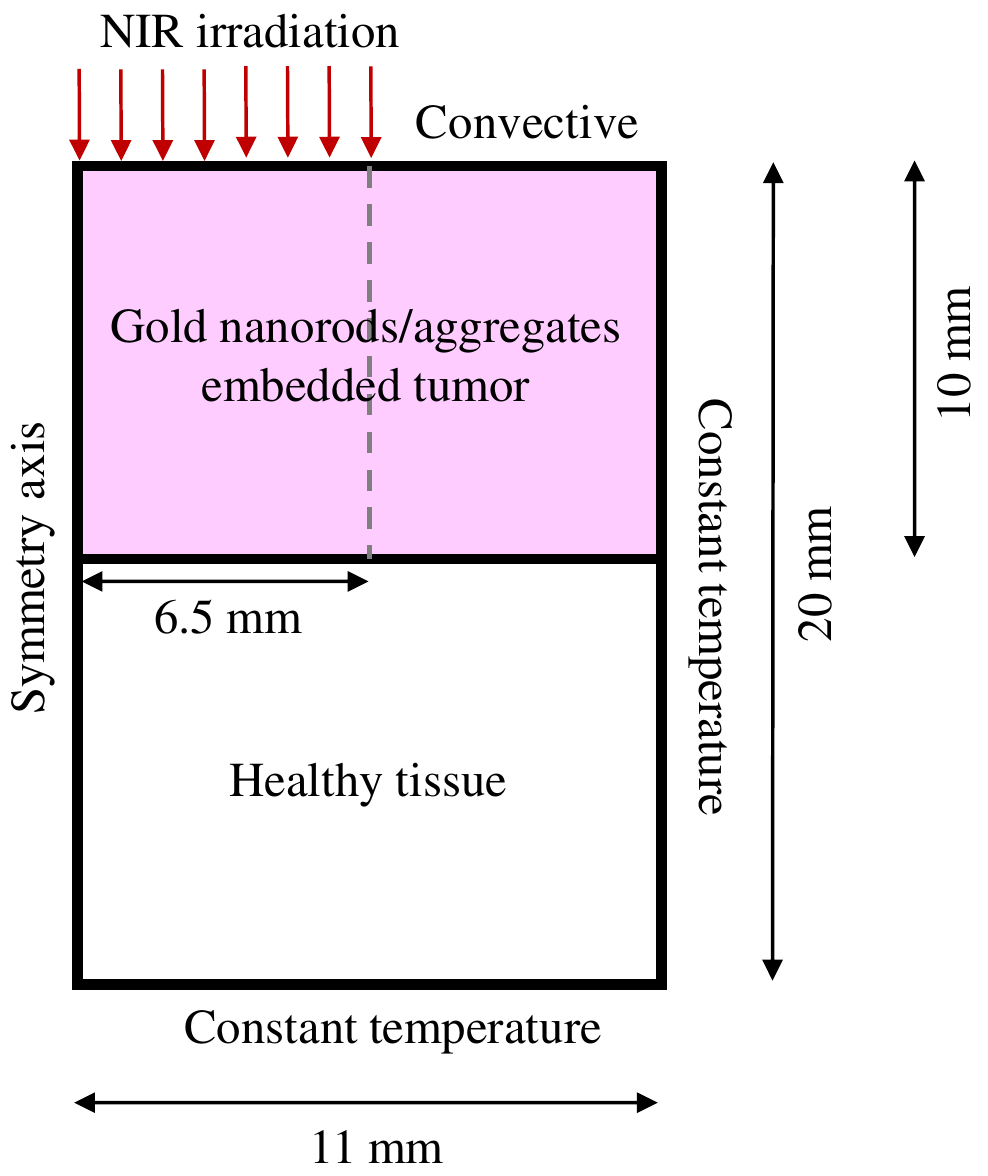}
\caption{Schematic of two-dimensional axis-symmetric computational modeling shows the different domains and boundary conditions.}
\label{fig:computational_domain}
\end{figure}

Figure~\ref{fig:computational_domain} shows the schematic of the computational modelling domain of the considered gel phantom. We solved Penne's Bioheat equation~\cite{wissler1998pennes}. The photothermal transport phenomenon is validated through numerical simulations. For which the same cylindrical agarose gel phantom size of diameter $22\,mm$ and depth $10\,mm$ located above healthy tissue of diameter and depth is $22\,mm$ and $10\,mm$ was considered for numerical simulation. Further, to reduce the computation cost, the three-dimensional cylindrical numerical model was reduced to the two-dimensional axis-symmetric model having radius and depth are $11\,mm$ and $20\,mm$ as Fig.~~\ref{fig:computational_domain}. Computation was carried out of the two-dimensional axis-symmetric model using the finite difference method (FDM).  The tumor-tissue domain was kept at a constant temperature at the right and bottom boundary to mimic the constant body temperature. The left boundary was symmetric due to the axial symmetric condition. The top surface was kept as a convective boundary as it is exposed to the environment. NIR irradiation was on the tumor top surface using $12\,mm$ diameter fiber bundle. The beam diameter was measured $13\,mm$ at the tumor top surface. Hence, for the two-dimensional domain, only $6.5\,mm$ radius was exposed to NIR radiation Fig.~\ref{fig:computational_domain}. The other tumor region remains unirradiated.

During plasmonic photothermal therapy, NIR radiation is absorbed by the plasmonic nanoparticles  present in the tumor which gives rise to the temperature of the tumor-tissue domain which eventually results in thermal damage. Here, light-tissue interaction is solved by Beer-Lambert's equation to get energy absorption and thermal transport is solved by Pennes' bioheat equation with blood perfusion term set to zero. Numerical simulation was performed in MATLAB using the FDM algorithm for which the space was discretized in $34\times61$ nodes.
Generally, bare tumor/tissue exposed to NIR irradiation causes negligible energy absorption due to a low absorption coefficient. However, gold nanorods or aggregates embedded tumor absorbs NIR irradiation significantly due to a higher absorption coefficient. For such media, discretized Beer-Lambert's equation for single wavelength shown in Eq. (\ref{Beer eq}) is solved to obtain the attenuated intensity when NIR radiation propagates through a tumor layer of thickness $\Delta z$. 
\begin{equation}\label{Beer eq}
    I_{\Delta z}(\lambda)=I_0(\lambda)e^{-(\mu_a(\lambda)+\mu_s(\lambda))\Delta z}
\end{equation}
where $I_{\Delta z}(\lambda)$ is attenuated intensity of NIR radiation travelling through tissue layer $\Delta z$ at particular wavelength $\lambda$, $I_0(\lambda)$ is incident radiation at particular wavelength $\lambda$, $\mu_a(\lambda)$ is the spectral absorption coefficient of the gold nanorods/aggregates and $\mu_a(\lambda)$ is the spectral scattering coefficient of the gold nanorods/aggregates. Since a broadband NIR source was used for the photothermal therapy. Hence, Beer-Lambert's equation is solved for each wavelength. To make computation simple, it was assumed that the incident NIR irradiation has no spectral variation in intensity i.e., each wavelength possesses the same intensity. The total incident intensity and total attenuated intensity by a tumor layer of $\Delta z$ is obtained by summing all intensities within the whole broadband spectral range discretized into 1~nm wavelength intervals which are mathematically expressed as \[I_{0} = \sum_{i=n_1}^{n_2}{I_{0}(\lambda_i)}\] \[I_{\Delta z} = \sum_{i=n_1}^{n_2}{I_{\Delta z}(\lambda_i)}\]\\
Where $n_1$ and $n_2$ are starting and ending wavelengths of broadband NIR irradiation.\\\\
Further, total internal heat generation $Q_{gen}$ within gold nanorods/aggregates embedded tumor layer of thickness $\Delta z$ is calculated using Equation \ref{Qgen}.

\begin{equation}\label{Qgen}
    Q_{gen}(\Delta z) = \frac{I_0-I_{\Delta z}}{\Delta z}
\end{equation}
%
To solve the temperature distribution within tumor-tissue domain, Pennes' bioheat equation is solved using the 
finite difference method. Equation \ref{FDM} shows the discretized form of Pennes' bioheat equation in two-dimension which is implemented in MATLAB.
\begin{align}\label{FDM}
    \frac{\rho C (T_{m,n}^{p+1}-T_{m,n}^p)}{\Delta t} = &\kappa\frac{T_{m+1,n}^{p}+T_{m-1,n}^{p}-2T_{m,n}^p}{\Delta x^2}+\kappa\frac{T_{m,n+1}^{p}+T_{m,n-1}^{p}-2T_{m,n}^p}{\Delta y^2} \nonumber\\
    &+\omega \rho_b C_b(T_b -T_{m,n}^p)+Q_{met}+Q_{gen}
\end{align}
where $\rho ,\,C ,\,\kappa ,\,\omega ,\, \textrm{and}\,Q_{met}$ are density, specific heat capacity, thermal conductivity, blood perfusion and metabolic heat generation of tumor/tissue. $\rho_b,\,C_b,$ and $T_b$ are the density, specific heat capacity and temperature of the blood. $Q_{gen}$ is the internal heat generation computed from Beer-Lambert's equation. The list of parameters and its values used to solve the Eq.~(\ref{FDM}) are listed in Table~\ref{tab:para_PBE}. $\Delta t,\, \Delta x$ and $\Delta y$ are time steps, spatial steps in the \textit{x}-direction and spatial step in the \textit{y}-direction. Superscript \textit{p} denotes the instantaneous time step and subscripts \textit{m} and \textit{n} denote node numbers along the \textit{x} and \textit{y} direction respectively. The numerical domain was divided into $34\times 61$ and the time step of 0.1\, second was considered.

\begin{table}[h]
\centering
\caption{List of parameters and its values used for solving the Penne's bioheat equation. }
\begin{tabular}{| l | c | l | c |}
\hline 
Parameters & Values & Parameters & Values \\ 
\hline \hline
$\rho$      & 1000~kg/m$^3$		& $\rho_b$		& 1000~kg/m$^3$  \\ \hline  
$C$ 		& 4200~J/kg/K   	& $C_b$ 		& 4200~J/kg/K	\\ \hline  
$\kappa$    & 0.5~W/m/K     	& $T_b$ 		& 23~$^{\circ}$C \\ \hline 
$\omega$ 	& 0 1/s  	& $Q_{met}$ 	& 0 W/m$^3$	\\ \hline  
\end{tabular}
\label{tab:para_PBE}
\end{table}

\begin{figure}[h]
\centering
\includegraphics[width=1\textwidth]{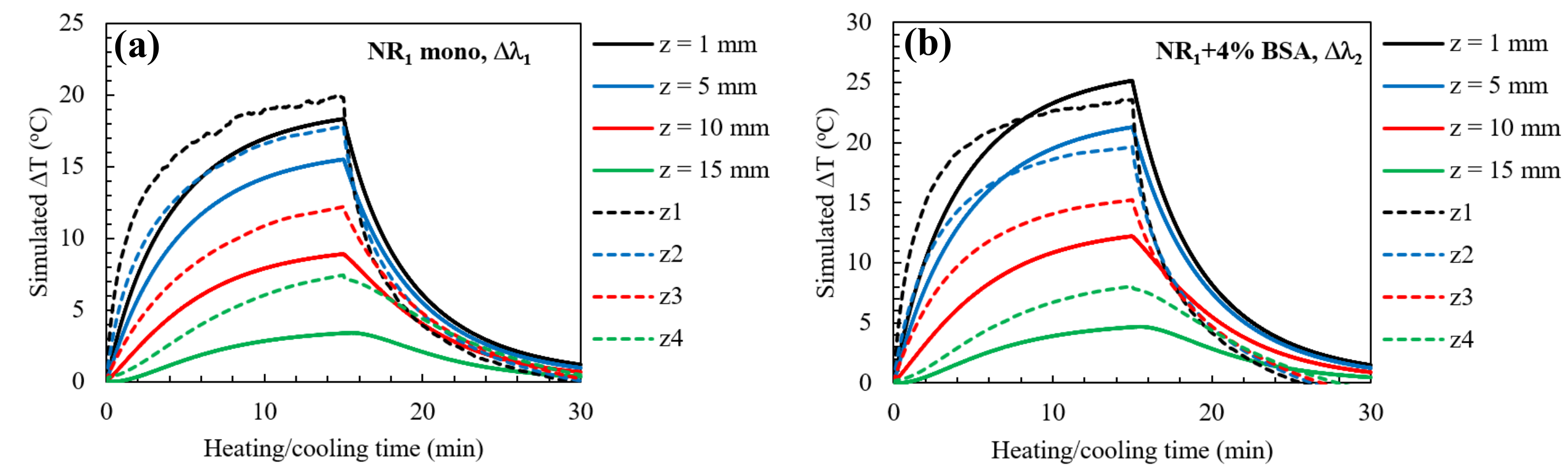}
\caption{Simulated temperature rise ($^{\circ}$C) after irradiating the tissue gel phantom embedded with gold nanorod and their aggregates (a) NR1 monodispersive and (b) NR1 aggregates with 4\%BSA. The dotted curves are experimental for comparison.}
\label{fig:NR1_pptt_sim}
\end{figure}

\begin{figure}[h]
\centering
\includegraphics[width=1\textwidth]{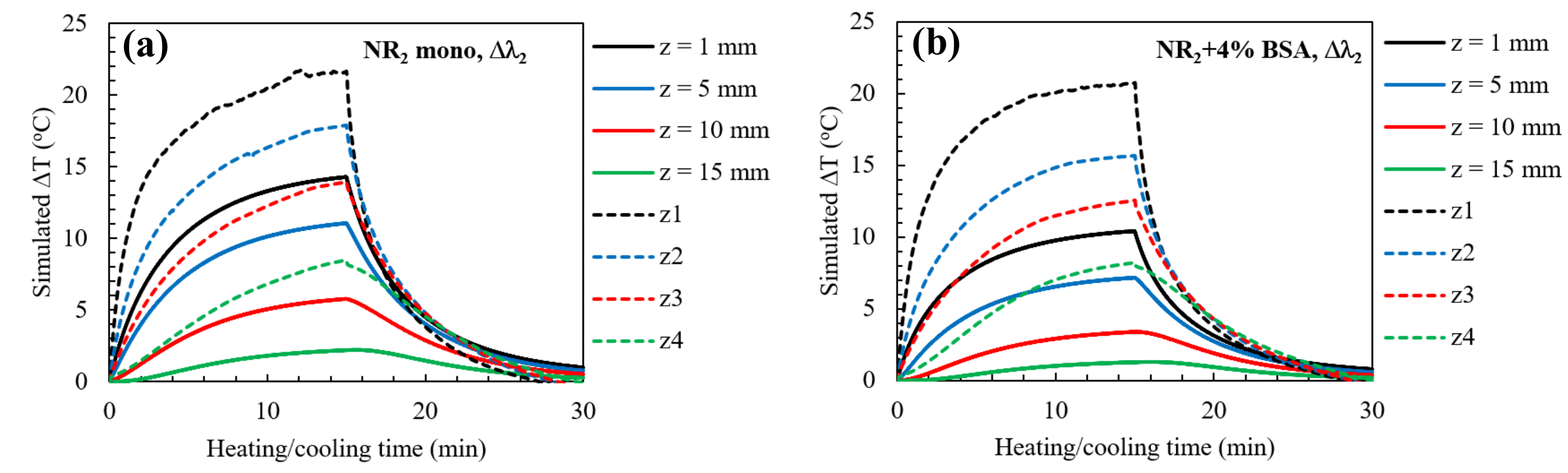}
\caption{Simulated temperature rise ($^{\circ}$C) after irradiating the tissue gel phantom embedded with gold nanorod and their aggregates (a) NR2 monodispersive and (b) NR2 aggregates with 4\%BSA. The dotted curves are experimental for comparison.}
\label{fig:NR2_pptt_sim}
\end{figure}
The numerically calculated temperature rise for the first set of samples is shown in Fig.~\ref{fig:NR1_pptt_sim}. The dotted curves are experimental to compare with simulation results. After the 15 minutes of irradiation time, temperature rise in the nanorods embedded gel phantom at the depth of 1~mm, 5~mm, 10~mm and 15~mm were 18.33~$^{\circ}$C, 15.51~$^{\circ}$C, 8.91~$^{\circ}$C, and 3.39~$^{\circ}$C respectively while for the aggregates embedded case were 25.16~$^{\circ}$C, 21.29~$^{\circ}$C,12.22~$^{\circ}$C, 4.64~$^{\circ}$C respectively. We can see that due to the aggregation there is more temperature rise as was observed experimentally. There is good agreement between the experiment and simulation for the samples synthesised by method-1. The calculated temperature rise for the second set of samples is shown in Fig.~\ref{fig:NR2_pptt_sim}. To compare, the experimental data are also plotted as dotted lines. The temperature rise after the 15~minutes irradiation time at depth of 1~mm, 5~mm, 10~mm and 15~mm were 14.29~$^{\circ}$C, 11.06~$^{\circ}$C, 5.78~$^{\circ}$C, and 2.19~$^{\circ}$C respectively while for the aggregates case these values were 10.43~$^{\circ}$C, 7.18~$^{\circ}$C, 3.40~$^{\circ}$C and 1.29~$^{\circ}$C respectively. In the latter case, there is some variation between experimental and simulation results, however, there is a similarity. The simulated result of the photothermal response of NR$_2$+4\%BSA is more deviating from the experimental than NR$_2$. This might be due to the reason that the LSPR resonance of these aggregates is out of the spectral range of bandpass filter-2 and scattering is quite large. Therefore, aggregates of smaller nanorods are better than the larger nanorods for plasmonic photothermal applications.

\section{Conclusion}
In conclusion, we have synthesized two types of gold nanorods and their aggregates to test the \textit{in vitro} PPTT measurement and to show similar trends using a broadband NIR light source. The aggregation also depends on the type of nanoparticles, protocols, and constitutive chemicals used to synthesize these. The use of the aggregated form of the nanoparticles is preferable for therapeutic applications to get better performance in photothermal heat generation. By choosing the appropriate size and shape of the nanoparticles one can also tune the LSPR in the second biological therapeutic window easily. We have carried out the PPTT testing embedding the synthesized monodispersive gold nanorods or their aggregates in the gel phantom prepared using agarose powder. The photothermal measurement showed that it is better to use aggregates instead of monodispersive nanoparticles for the PPTT.  In many situations, it is better to use of broadband light source instead of single-wavelength laser light. The use of broadband light would address the slight changes in the plasmonic resonances of the nanoparticles or even their aggregates. The present study is useful for the better performance of PPTT in tumor treatment, through use of aggregates of small sized nanoparticles for opting to the second biological window for deeper penetration of light in tissue and enhanced photothermal response. 

\section*{Acknowledgements}
D.P. acknowledges the CSIR-India for the Nehru Science Postdoctoral Research Fellowship number HRDG/CSIR-Nehru PDF/EN, ES \& PS/EMR-I/04/2019. D.P. and S.S. acknowledge the support of the CSIR-CSIO Chandigarh for hosting the research.
\bibliographystyle{unsrtnat}
\bibliography{references}  






\end{document}